\documentclass[12pt]{article}
\usepackage{amssymb}
\usepackage{graphicx}
\newtheorem{theorem}{Theorem}
\newtheorem{theoremMy}{Theorem}
\newtheorem{definition}{Definition}

\begin{document}
\pagestyle{empty}

\title{On Solving Travelling Salesman Problem with Vertex
Requisitions\thanks{This research of the authors is supported by
the  Russian Science Foundation Grant (project no.~15-11-10009).}
}


\author{Anton V. EREMEEV\\ \it{Omsk Branch of Sobolev Institute of
Mathematics SB RAS,}\\
\it{Omsk State University n.a.~F.M.~Dostoevsky}\\
\it{~eremeev@ofim.oscsbras.ru}
       \and
        Yulia V. KOVALENKO \\ \it{Sobolev Institute of
Mathematics SB RAS,}\\
\it{~julia.kovalenko.ya@yandex.ru} }

\date{}
\maketitle

\begin{center}
{ Received:  / Accepted: }
\end{center}

\maketitle

\noindent {\bf Abstract:} We consider the Travelling Salesman
Problem with Vertex Requisitions, where for each position of the tour
 at most two possible vertices are given. It is
known that the problem is strongly NP-hard. The proposed algorithm for this
problem has less time complexity compared to the previously known one.
In particular, almost all feasible instances of the problem are
solvable {in~$O(n)$ time} using the new algorithm, where $n$ is the number of vertices.
The developed approach also helps in fast enumeration of a neighborhood in the
local search and yields an integer programming model with~$O(n)$
binary variables for the problem.\\



\section{INTRODUCTION} \label{intro}

\noindent  \label{sec:Introduction} The {\sc Travelling Salesman
Problem}~(TSP) is one of the well-known NP-hard combinatorial
optimization problems~\cite{GJ}: given a complete arc-weighted
digraph with~$n$ vertices, find a shortest travelling salesman
tour (Hamiltonian circuit) in it.

The {\sc TSP with Vertex Requisitions}~(TSPVR) was formulated
by\linebreak A.I.~Serdyukov in~\cite{SAI1985}: find a shortest
travelling salesman tour, passing at $i$-th position a vertex from
a given subset~$X^i,\ i=1,\dots,n$. A special case where
$|X^i|=n,\ i=1,\dots,n,$ is equivalent to the TSP.

This problem can be interpreted in terms of scheduling theory.
Consider a single machine that may perform a set of
operations~$X=\{x_1,\dots,x_n\}$. Each of the identical jobs
requires processing all~$n$ operations in such a sequence that
the $i$-th operation  belongs to a given subset~$X^i\subseteq X$ for all ${i=1,\dots,n}$.
A  setup time is needed to switch the
machine from one operation of the sequence to another.  Moreover,
after execution of the last operation of the sequence the machine
requires a changeover to the first operation of the sequence to
start processing of the next job.  The problem is to find a
feasible sequence of operations, minimizing the cycle time.

{\sc TSP with Vertex Requisitions} where $|X^i|\leq k$,
${i=1,\dots,n,}$ was called $k$-{\sc TSP with Vertex Requisitions}
($k$-TSPVR) in~\cite{SAI1985}. The complexity of $k$-TSPVR  was
studied in~\cite{SAI1985} for different values of~$k$ on graphs
with small vertex degrees. In~\cite{SAI} A.I.~Serdyukov proved the
NP-hardness of $2$-TSPVR in the case of complete graph and showed
that almost all feasible instances of the problem are solvable in $O(n^2)$
time. In this paper, we propose an algorithm for
$2$-TSPVR with {time complexity~$O(n)$} for almost all
feasible problem instances. The developed approach also has some
applications to local search and integer programming formulation
of $2$-TSPVR.

The paper has the following structure. In
Section~\ref{sec:statement}, a formal definition of $2$-TSPVR is
given. In Section~\ref{sec:algorithm}, an algorithm for
this problem is presented. In Section~\ref{sec:polinom}, a
modification of the algorithm is proposed with an improved time
complexity and it is shown that almost all feasible instances of the
problem are solvable in {time~$O(n)$.} In
Section~\ref{sec:locsech}, the developed approach is used to
formulate and  enumerate efficiently a neighborhood for local
search. In Section~\ref{sec:mip}, this approach allows to
formulate an integer programming model for $2$-TSPVR using~$O(n)$
binary variables. The last section contains the concluding
remarks.

\section{PROBLEM FORMULATION AND ITS HARDNESS} \label{sec:statement}

\noindent $2$-{\sc TSP with Vertex Requisitions} is formulated as
follows. Let ${G=(X,U)}$ be a complete arc-weighted
digraph, where ${X=\{x_1,\dots,x_n\}}$ is the set of vertices,
${U=\{(x,y):\ x,y \in X, x\ne y\}}$ is the set of arcs with
non-negative arc weights $\rho(x,y),$ ${(x,y)\in U}$. Besides
that, a system of vertex subsets (requisitions) $X^i\subseteq X,\
i=1,\dots,n,$ is given, such that $1 \leq |X^i|\leq 2$ for all
${i=1,\dots,n}$.

Let $F$ denote the set of bijections from ${X_n:=\{1,\dots,n\}}$
to~$X$, such that $f(i)\in X^i,\ i=1,\dots,n,$ for all~$f\in F$.
The problem consists in finding such a mapping ${f}^*\in F$ that
${\rho}({f}^*)=\min\limits_{f\in F} {\rho}(f)$, where
${\rho}({f})=\sum\limits_{i=1}^{n-1}
\rho({f}(i),{f}(i+1))+\rho({f}(n),{f}(1))$ for all $f\in F$.
Later on the symbol~$I$ is used for the instances of this problem.

Any feasible solution uses only the arcs that start in a
subset~$X^i$ and end in $X^{i+1}$ for some~$i\in \{1,\dots,n\}$
(we assume $n+1:=1$). Other arcs are irrelevant to the
problem and we assume that they are not given in a problem input~$I$.

2-TSPVR is strongly NP-hard~\cite{SAI}. The proof of this fact
in~\cite{SAI} is based on a reduction of {\sc Clique} problem to a
family of instances of $2$-TSPVR with integer input data, bounded
by a polynomial in problem length. Therefore, in view of
sufficient condition for non-existence of Fully Polynomial-Time
Approximation Scheme~(FPTAS) for strongly NP-hard
problems~\cite{GJ1978}, the result from~\cite{SAI} implies that
$2$-TSPVR does not admit an FPTAS, provided that P$\ne$NP. The
$k$-TSPVR with $k \ge 3$ cannot be approximated with any constant
or polynomial factor of the optimum in polynomial time, unless
P=NP, as follows from~\cite{SAI79}.

\section{SOLUTION METHOD}
\label{sec:algorithm}

\noindent Following the approach of A.I.~Serdyukov~\cite{SAI}, let
us consider a bipartite graph $\bar{G}=(X_{n},X,\bar{U})$ where
the two subsets of vertices of bipartition $X_n, X$ have equal
sizes and the set of edges is ${\bar{U}=\{\{i,x\}:}$ ${i\in
X_{n},\ x\in X^i}\}$. Now there is a one-to-one correspondence
between the set of perfect matchings~$\mathcal{W}$ in the
graph~$\bar{G}$ and the set~$F$ of feasible solutions to a problem
instance~${I}$: Given a perfect matching~$W \in \mathcal{W}$ of
the form $\{\{1,x^1\},\{2,x^2\},\dots,\{n,x^n\}\}$, this mapping
produces the tour $(x^1,x^2,\dots,x^n)$.

An edge~$\{i,x\}\in \bar{U}$ is called {\it special} if $\{i,x\}$
belongs to all perfect matchings in the graph~$\bar{G}$.
Let us also call the vertices of the graph~$\bar{G}$ {\it special}, if
they are incident with special edges.

Supposing that~$\bar{G}$ is given by the lists of adjacent vertices, the
special edges and edges that do not belong to any
perfect matching in the graph~$\bar{G}$ may be efficiently
computed by the Algorithm~1 described below.
After that all edges, except for the special
edges and those adjacent to them, are slit into cycles. Note that the method of finding all special edges and cycles in
the graph~$\bar{G}$ was not discussed in~\cite{SAI}.\\

{\bf \textit{Algorithm 1. Finding special edges in the graph~$\bar{G}$}}\\

{\bf Step~1} {\it(Initialization)}{\bf.} Assign
$\bar{G}':=\bar{G}$.

{\bf Step~2.} Repeat Steps 2.1-2.2 while it is possible:

{\bf Step~2.1} {\it(Solvability test)}{\bf}. If the graph~$\bar{G}'$ contains a vertex of degree~0 then problem~${I}$
is infeasible, terminate.

{\bf  Step~2.2} {\it(Finding a special edge)}{\bf.} If the graph~$\bar{G}'$ contains a vertex~$z$ of degree~1, then store
the corresponding edge~$\{z,y\}$ as a special edge and remove its endpoints $y$ and $z$ from~$\bar{G}'$.\\

Each edge of the graph~$\bar{G}$ is visited and deleted
at most once (which takes~$O(1)$ time). The number of edges
$|\bar{U}|\leq 2n$. So the time complexity of Algorithm~1 is~$O(n)$.

Algorithm~1 identifies the case when problem~${I}$
is infeasible. Further we consider only feasible instances of 2-TSPVR and bipartite graphs corresponding to them.

After the described preprocessing the resulting graph~$\bar{G}'$
is $2$-regular (the degree of each vertex equals~2) and its
components are even cycles. The cycles of the graph~$\bar{G}'$ can be
computed in~$O(n)$ time using the Depth-First Search algorithm
(see e.g.~\cite{KLR}). Note that there are no other edges in the
perfect matchings of the graph~$\bar{G}$, except for the special edges
and edges of the cycles in $\bar{G}'$. In what follows,
$q(\bar{G})$~denotes the number of cycles in the graph~$\bar{G}$ (and
in the corresponding graph~$\bar{G}'$).

\begin{figure}[h]
\begin{center}
\includegraphics[height=8cm,width=6.9cm]{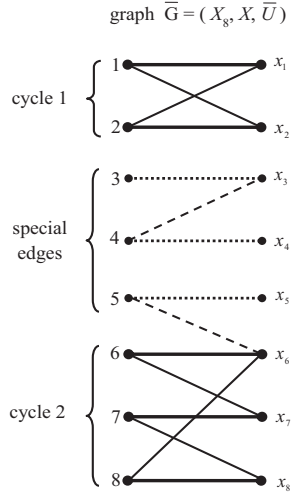}\\
\caption{An instance~$I$ with $n=8$ and system of vertex
requisitions: ${X^1=\{x_1,x_2\}}$, $X^2=\{x_1,x_2\}$,
$X^3=\{x_3\}$, $X^4=\{x_3,x_4\}$, $X^5=\{x_5,x_6\}$,
$X^6=\{x_6,x_7\}$, $X^7=\{x_7,x_8\}$, $X^8=\{x_6,x_8\}$. Here the
edges drawn in bold define one maximal matching of a cycle, and
the rest of the edges in the cycle define another one. The special
edges are depicted by dotted lines. The edges depicted by dashed
lines do not belong to any perfect matching. The feasible
solutions of the instance are
$f^1=(x_1,x_2,x_3,x_4,x_5,x_6,x_7,x_8),\
f^2=(x_1,x_2,x_3,x_4,x_5,x_7,x_8,x_6),$
${f^3=(x_2,x_1,x_3,x_4,x_5,x_6,x_7,x_8),}$
$f^4=(x_2,x_1,x_3,x_4,x_5,x_7,x_8,x_6)$.} \label{fig:bijection}
\end{center}
\end{figure}

Each cycle~${j,\ j=1,\dots,q(\bar{G})}$,
contains exactly two maximal (edge disjoint) perfect matchings, so
it does not contain any special edges. Every perfect matching
in~$\bar{G}$ is uniquely defined by a combination of maximal
matchings chosen in each of the cycles and the set of all special
edges (see Fig.~\ref{fig:bijection}). Therefore, 2-TSPVR is
solvable by the following
algorithm.\\

{\bf \textit{Algorithm 2. Solving 2-TSPVR}}\\

{\bf Step~1.}~Build the bipartite graph~$\bar{G}$, identify the
set of special edges and cycles and find all maximal matchings in
cycles.

{\bf Step~2.}~Enumerate all perfect matchings~$W\in \mathcal{W}$
of $\bar{G}$ by combining the maximal matchings of cycles
and joining them with special edges.

{\bf Step~3.}~Assign the corresponding solution~$f\in F$ to
each~$W\in \mathcal{W}$ and compute~$\rho(f)$.

{\bf Step~4.}~Output the result~$f^*\in F$, such that
$\rho(f^*)=\min\limits_{f\in F}\rho(f)$.

To evaluate the Algorithm~2, first note that maximal matchings in
cycles are found easily in $O(n)$ time. Now
$|F|=|\mathcal{W}|=2^{q(\bar{G})}$ so the time complexity of
Algorithm~2 of solving 2-TSPVR is $O(n 2^{q(\bar{G})})$, where
${q(\bar{G})\leq \lfloor \frac{n}{2}\rfloor}$ and the last
inequality is tight.\\

\section{IMPROVED ALGORITHM}
\label{sec:polinom}

\noindent In~\cite{SAI}, it was shown that almost all feasible instances of
2-TSPVR have not more than~$n$ feasible solutions and may be
solved in quadratic time. To describe this result precisely, let
us give the following

\begin{definition}{\rm~\cite{SAI}}
\label{def:good_graph} A graph~$\bar{G}=(X_n,X,\bar{U})$ is called
``good'' if it satisfies the inequality ${q(\bar{G})\leq 1.1~{\rm
ln}(n)}$.
\end{definition}

Note that any problem instance~$I$,  which corresponds to a ``good''\
graph~$\bar{G}$, has at most {$2^{1.1 \ln
n}<n^{0.77}$ feasible solutions.}

Let $\bar{\mathcal{\chi}}_n$ denote the set of ``good'' bipartite
graphs~$\bar{G}=(X_n,X,\bar{U})$ , and let ${\mathcal{\chi}}_n$ be
the set of {\em all}\ bipartite graphs~$\bar{G}=(X_n,X,\bar{U})$.
The results of A.I.~Serdyukov from~\cite{SAI} imply

\begin{theorem}
$|\bar{\mathcal{\chi}}_n|/|{\mathcal{\chi}}_n| \longrightarrow 1$ as
$n\to \infty$.\label{th4}
\end{theorem}

The proof~of Theorem~\ref{th4} from~\cite{SAI} is provided in the
appendix for the sake of completeness. According to the frequently
used terminology (see e.g. \cite{Chvatal}), this theorem means
that almost all feasible instances~$I$ have at
most~$n^{0.77}$ feasible solutions and thus they are solvable in
$O(n^{1.77})$ time {by Algorithm~2}.

Using the approach from~\cite{EK2014II} we will now modify
Algorithm~2 for solving 2-TSPVR in
$O\left(q(\bar{G})2^{q(\bar{G})}+n\right)$ time. Let us carry out
some preliminary computations before enumerating all possible
combinations of maximal matchings in cycles in order to speed up
the evaluation of objective function. We will call a {\em contact
between cycle~$j$ and  cycle~$j'\ne j$ (or between cycle~$j$ and a
special edge)} the pair of vertices~$(i,i+1)$ (we assume $n+1:=1$) in the
left-hand part of the graph~$\bar{G}$, such that one of the vertices
belongs to the cycle~$j$ and the other one belongs to the cycle~$j'$
(or the special edge). A {\em contact inside a cycle} will mean a
pair of vertices in the left-hand part of a cycle, if their
indices differ exactly by one, or these vertices are~$(n,1)$.

Consider a cycle~${j}$. If a contact~$(i,i+1)$  is
present inside this cycle, then each of the two maximal
matchings~$w^{0,j}$ and $w^{1,j}$ in this cycle determines the $i$-th arc of a tour in the graph~$G$.
Also, if the cycle~$j$ has a contact~$(i,i+1)$ to a special
edge, each of the two maximal matchings~$w^{0,j}$ and $w^{1,j}$
also determines the $i$-th arc of a tour in the graph~$G$.
For each of the matchings~${w^{k,j}, \ k=0,1}$, let the
sum of the weights of arcs determined by the contacts inside
the cycle~$j$ and the contacts to special edges be denoted
by~$P^{k}_j$.

If cycle~${j}$ contacts to cycle~${j', \ j'\ne j}$, then each
combination of the maximal matchings of these cycles determines the $i$-th arc of a tour in the graph~$G$
for any contact~$(i,i+1)$
between the cycles. If a maximal matching is chosen in each of the
cycles, one can sum up the weights of the arcs in~$G$
determined by  all contacts between cycles~${j}$ and ${j'}$.
This yields four values which we denote by $P^{(0,0)}_{j j'}$,
$P^{(0,1)}_{j j'}$, $P^{(1,0)}_{j j'}$ and $P^{(1,1)}_{j j'}$,
where the superscripts identify the matchings chosen in each of
the cycles~$j$ and $j'$ respectively.

Parameters $P^{0}_j$, $P^{1}_j$, $P^{(0,0)}_{j j'}$, $P^{(0,1)}_{j j'}$, $P^{(1,0)}_{j
j'}$ and $P^{(1,1)}_{j j'}$ can be found as follows. Suppose that
intermediate values of $P^{0}_j$, $P^{1}_j$ for $j=1,\dots,q(\bar{G})$ are stored in
one-dimensional arrays of size~$q(\bar{G})$, and
intermediate values of $P^{(0,0)}_{j j'}$, $P^{(0,1)}_{j j'}$,
$P^{(1,0)}_{j j'}$ and $P^{(1,1)}_{j j'}$ for $j,\
j'=1,\dots,q(\bar{G})$ are stored in two-dimensional arrays of
size $q(\bar{G})\times q(\bar{G})$.
Initially, all of these values are assumed to be zero
and they are computed in an iterative way by the consecutive
enumeration of pairs of vertices~$(i,i+1)$, ${i=1,\dots,n-1}$, and $(n,1)$
in the left-hand part of the graph~$\bar{G}$. When we consider
a pair of vertices~$(i,i+1)$ or $(n,1)$, at most four parameters
(partial sums) are updated depending on whether the vertices
belong to different cycles or to the same cycle, or one of the
vertices is special. So the overall time complexity of the
pre-processing procedure is~$O(q^2(\bar{G})+n)$.

Now all possible combinations of the maximal matchings in cycles
may be enumerated using a Grey code (see e.g.~\cite{RND}) so that
the next combination differs from the previous one by altering a
maximal matching only in one of the cycles. Let the binary
vector~$\delta=(\delta_1,\dots,\delta_{q(\bar{G})})$ define
assignments of the maximal matchings in cycles. Namely,
$\delta_j=0$, if the matching~$w^{0,j}$ is chosen in the cycle~$j$;
otherwise (if the matching~$w^{1,j}$ is chosen in the cycle~$j$), we
have~$\delta_j=1$. This way every vector~$\delta$ is bijectively
mapped into a feasible solution~$f_{\delta}$ to 2-TSPVR.

In the process of enumeration, a step from the current
vector~$\bar{\delta}$ to the next vector~$\delta$ changes the
maximal matching in one of the cycles~$j$. The new value of
objective function~$\rho(f_{\delta})$ may be computed via the
current value~$\rho(f_{\bar{\delta}})$ by the formula
$\rho(f_{\delta})=\rho(f_{\bar{\delta}})-P^{{\bar
\delta}_j}_j+P^{{\delta}_j}_j-\sum\limits_{j'\in A(j)}P^{({\bar
\delta}_j,{\bar \delta}_{j'})}_{jj'}+\sum\limits_{j'\in
A(j)}P^{({\delta}_j,{\delta}_{j'})}_{jj'}$, where $A(j)$ is the
set of cycles contacting to the cycle~$j$. Obviously, $|A(j)|\leq
q(\bar{G})$, so updating the objective function value for the next
solution requires~$O(q(\bar{G}))$ time, and the overall time
complexity of the modified algorithm for solving 2-TSPVR
is~$O\left(q(\bar{G})2^{q(\bar{G})}+n\right)$.

In view of Theorem~\ref{th4} we conclude that using this
modification of Algorithm~2 almost all feasible instances of 2-TSPVR are
solvable in $O(n^{0.77} \ln n + n)=O(n)$ time.

\section{LOCAL SEARCH}
\label{sec:locsech}

\noindent A local search algorithm starts from an initial feasible
solution. It moves iteratively from one solution to a better
neighboring solution and terminates at a local optimum. The number
of steps of the algorithm, the time complexity of one step, and
the value of the local optimum depend essentially on the
neighborhood. Note that neighborhoods, often used for the
classical TSP (e.g. k-Opt, city-swap,
Lin-Kernighan~\cite{Kochetov2008}), will contain many infeasible
neighboring solutions if applied to 2-TSPVR  because of the vertex
requisition constraints.

A local search method with a specific neighborhood for
2-TSPVR may be constructed using the relationship between the
perfect matchings in the graph~$\bar{G}$ and the feasible solutions.
The main idea of the algorithm consists in building a neighborhood
of a feasible solution to 2-TSPVR on the basis of a Flip
neighborhood of the perfect matching, represented by the maximal
matchings in cycles and the special edges.

Let the binary vector
$\delta=(\delta_1,\dots,\delta_{q(\bar{G})})$ denote the
assignment of the maximal matchings to cycles as above. The set
of~$2^{q(\bar{G})}$ vectors~$\delta$ corresponds to the set of
feasible solutions by a one-to-one mapping~$f_{\delta}$. We assume
that a solution $f_{\delta'}$ belongs to the {\em Exchange}
neighborhood of solution~$f_{\delta}$ iff the vector~$\delta'$ is
within Hamming distance~1 from~$\delta$, i.e. $\delta'$ belongs to
the Flip neighborhood of vector~$\delta$.

Enumeration of the Exchange neighborhood takes~$O(q^2(\bar{G}))$
time if the preprocessing described in Section~\ref{sec:polinom}
is carried out before the start of the local search (without the
preprocessing it takes $O(nq(\bar{G}))$ operations). Therefore,
for almost all feasible instances~$I$, the Exchange neighborhood may be
enumerated in $O({\rm ln}^2(n))$ time.

\section{MIXED INTEGER LINEAR PROGRAMMING MODEL}
\label{sec:mip}

\noindent The one-to-one mapping between the maximal matchings in
cycles of the graph~$\bar{G}$ and feasible solutions to 2-TSPVR may be
also exploited in formulation of a mixed integer linear
programming model.

Recall that $P^{0}_j$ ($P^1_j$) is the sum of weights of all arcs
of the graph~$G$ determined by  the contacts inside the cycle~$j$ and
the contacts of the cycle~$j$ with special edges, when the maximal
matching~$w^{0,j}$ ($w^{1,j}$) is chosen in the cycle~$j$,
${j=1,\dots,q(\bar{G})}$. Furthermore, $P^{(k,l)}_{j j'}$ is the
sum of weights of arcs in the graph~$G$ determined by
the contacts between cycles~${j}$ and ${j'}$, if
the maximal matchings~$w^{k,j}$ and $w^{l,j'}$ are chosen in the cycles $j$ and $j'$ respectively, $k,l=0,1$, ${j=1,\dots,q(\bar{G})-1}$,
${j'=j+1,\dots,q(\bar{G})}$.
These values are computable in $O(n+q^2(\bar{G}))$ time as shown in Section~\ref{sec:polinom}.

Let us introduce the following Boolean variables:\\

{
\begin{tabular}{cll}
 $d_j=$ &
 $ \left\{
\begin{tabular}{l}
 $0,$ \mbox{ if matching~$w^{0,j}$}$ \mbox{ is chosen in the cycle}~j,$\\
 $1,$ \mbox{ if matching~$w^{1,j}$}$ \mbox{ is chosen in the cycle}~j,$\\
\end{tabular}  \right.
$
\end{tabular}
}\\
\vspace{0cm}\hspace{2.2cm}${j=1,\dots,q(\bar{G})}.$\\

The objective function combines the pre-computed arc weights for
all cycles, depending on the choice of matchings
in~$d=(d_1,\dots,d_{q(\bar{G})})$:

$$
\sum_{j=1}^{q(\bar{G})-1} \sum_{j'=j+1}^{q(\bar{G})}
\left(P^{(0,0)}_{j j'}(1-d_j)(1-d_{j'})+
                                   P^{(0,1)}_{j j'}(1-d_j)d_{j'}\right)
$$
\begin{equation}
 + \sum_{j=1}^{q(\bar{G})-1} \sum_{j'=j+1}^{q(\bar{G})}  \left(P^{(1,0)}_{j j'}d_j(1-d_{j'})+P^{(1,1)}_{j j'}d_jd_{j'}\right)
\label{criterion}
\end{equation}
$$
+\sum\limits_{j=1}^{q(\bar{G})}\left(P^{0}_j(1-d_j)+P^{1}_jd_j\right)
\to \min,
$$
\begin{equation}
d_j\in \{0,1\}, \ {j=1,\dots,q(\bar{G})}. \label{def}
\end{equation}

Let us define supplementary real variables in order to remove
non-linearity of the objective function: for
$k\in\{0,1\}$ we assume that $p_{j}^{(k)} \ge 0$ is an upper bound
on the sum of weights of arcs in the graph~$\bar{G}$ determined by
the contacts of the cycle~${j}$,
 if matching~$w^{k,j}$ is chosen in this cycle, i.e. $d_j=k$,
 ${j=1,\dots,q(\bar{G})-1}$.

Then the mixed integer linear programming model has the following
form:

\begin{equation}
\sum\limits_{j=1}^{q(\bar{G})-1}
\left(p^{(0)}_{j}+p^{(1)}_{j}\right)
+\sum_{j=1}^{q(\bar{G})}\left(P^{0}_j(1-d_j)+P^{1}_jd_j\right)\to
\min, \label{criterionLin}
\end{equation}
$$p^{(0)}_{j} \ge \sum\limits_{j'=j+1}^{q(\bar{G})} P^{(0,0)}_{j j'}\left(1-d_j-d_{j'}\right)+
\sum\limits_{j'=j+1}^{q(\bar{G})} P^{(0,1)}_{j j'}\left(d_{j'} -
d_j\right),$$
\begin{equation}
{j=1,\dots,q(\bar{G})-1}, \label{matching00_01}
\end{equation}
$$p^{(1)}_{j} \ge \sum\limits_{j'=j+1}^{q(\bar{G})}  P^{(1,0)}_{j j'}\left(d_j-d_{j'}\right)
+ \sum\limits_{j'=j+1}^{q(\bar{G})} P^{(1,1)}_{j
j'}\left(d_j+d_{j'}-1\right),$$
\begin{equation}
{j=1,\dots,q(\bar{G})-1}, \label{matching10_11}
\end{equation}
\begin{equation}
p^{(k)}_{j} \ge 0,\ k=0,1,\ {j=1,\dots,q(\bar{G})-1},
\label{defMatch}
\end{equation}
\begin{equation}
d_j\in \{0,1\}, \ {j=1,\dots,q(\bar{G})}. \label{def1}
\end{equation}

Note that if matching~$w^{0,j}$ is chosen for
the cycle~$j$ in an optimal solution of problem
(\ref{criterionLin})-(\ref{def1}), then inequality
(\ref{matching00_01}) holds for $p^{(0)}_{j}$  as equality  and
$p^{(1)}_{j}=0$. Analogously, if matching~$w^{1,j}$ is chosen for
the cycle~$j$, then inequality (\ref{matching10_11}) holds for
$p^{(1)}_{j}$ as equality and $p^{(0)}_{j}=0$. Therefore, problems
(\ref{criterion})-(\ref{def}) and
(\ref{criterionLin})-(\ref{def1}) are equivalent because a
feasible solution of one problem corresponds to a feasible
solution of another problem, and an optimal solution corresponds
to an optimal solution.

The number of real variables in model (\ref{criterionLin})-(\ref{def1}) is~$(2q(\bar{G})-2)$, 
the number of Boolean variables is~$q(\bar{G})$. The number of
constraints is $O(q(\bar{G}))$, where $q(\bar{G})\leq
\lfloor\frac{n}{2}\rfloor$. The proposed model may be used for
computing lower bound of objective function or in branch-and-bound
algorithms, even if the graph~$\bar{G}$ is not ``good''.

Note that there are a number of integer linear programming models
in the literature on the classical TSP, involving~$O(n^2)$ Boolean
variables. Model~(\ref{criterionLin})--(\ref{def1}) for 2-TSPVR
has at most~${\lfloor\frac{n}{2}\rfloor}$ Boolean variables and for
almost all feasible instances the number of Boolean variables is~$O({\rm
ln}(n))$.

\section{CONCLUSION}
\label{sec:conclusion}

\noindent We presented an algorithm for solving {\sc 2-TSP with
Vertex Requisitions}, that reduces the time complexity bound
formulated in~\cite{SAI}.
It is easy to see that the same approach
is applicable to the problem {\sc 2-Hamiltonian Path of Minimum Weight with
Vertex Requisitions}, that asks for a Hamiltonian Path of Minimum Weight
in the graph~$G$, assuming the same system of vertex requisitions as
in {\sc 2-TSP with Vertex Requisitions}.

Using the connection to perfect matchings in a supplementary
bipartite graph and some preprocessing we constructed a MIP model
with~$O(n)$ binary variables and a new efficiently searchable {\em
Exchange} neighborhood for problem under consideration.

Further research might address the existence of approximation
algorithms with constant approximation ratio for {\sc 2-TSP with
Vertex Requisitions}.



\begin{thebibliography}{}
%
%
\bibitem{GJ} Garey, M.R., Johnson, D.S.: Computers and intractability. A guide to the theory of
NP-completeness. W.H.~Freeman and Company, San Francisco, CA
(1979)

\bibitem{SAI1985}  Serdyukov, A.I.: Complexity of solving
the travelling salesman problem with requisitions on graphs with
small degree of vertices. Upravlaemye systemi. 26, 73--82 (1985)
(In Russian)

\bibitem{SAI} Serdyukov, A.I.: On travelling salesman problem with
prohibitions. Upravlaemye systemi. 17, 80--86 (1978) (In Russian)

\bibitem{GJ1978} Garey, M.R., Johnson, D.S.: Strong NP-completeness results:
Motivation, examples, and implications. Journal of the ACM. 25,
499--508 (1978)

\bibitem{SAI79} Serdyukov A.I.: On finding Hamilton
cycle (circuit) problem with prohibitions. Upravlaemye systemi.
19, 57--64 (1979) (In Russian)

\bibitem{KLR} Cormen, T.H., Leiserson, C.E., Rivest, R.L., Stein,
C.: Introduction to Algorithms, 2nd edition. MIT Press (2001)

\bibitem{Chvatal} Chvatal, V.: Probabilistic methods in graph
theory. Annals of Operations Research. 1, 171--182 (1984)

\bibitem{EK2014II} Eremeev, A., Kovalenko, J.: Optimal
recombination in genetic algorithms for combinatorial optimization
problems: Part II. Yugoslav Journal of Operations Research.
24~(2), 165-186 (2014)

\bibitem{RND} Reingold, E.M., Nievergelt, J., Deo, N.:
Combinatorial algorithms: Theory and Practice. Englewood Cliffs,
Prentice-Hall (1977)


\bibitem{Kochetov2008}
Kochetov, Yu. A.: Computational bounds for local search in
combinatorial optimization. Computational Mathematics and
Mathematical Physics. 48~(5), 747-763 (2008)

\bibitem{Feller70} Feller, W.:
An Introduction to Probability Theory and Its Applications. Vol.~1. John Wiley \& Sons,
New York, NY (1968)


\bibitem{Riordan58} Riordan, J.:
 An Introduction to Combinatorial Analysis.  John Wiley \& Sons,
New York, NY (1958)
\end{thebibliography}


\section*{APPENDIX}


Note that A.I.~Serdyukov in~\cite{SAI} used the term
{\em block} instead of term {\em cycle,} employed in
Section~\ref{sec:algorithm} of the present paper. A {\em block}
was defined in~\cite{SAI} as a maximal (by inclusion) 2-connected
subgraph of graph~$\bar{G}$ with at least two edges. However, in
each block of the graph~$\bar{G}$, the degree of every vertex equals~2
(otherwise $F= {\O}$ because the vertices of degree~1 do not
belong to blocks and the vertex degrees are at most~2 in the
right-hand part of~$\bar{G}$). So, the notions {\em block} and
{\em cycle} are equivalent in the case of considered bipartite
graph~$\bar{G}$. We use the term {\em block} in the proof~of
Theorem~\ref{th4} below, as in the original paper~\cite{SAI}, in
order to avoid a confusion of cycles in~$\bar{G}$ with cycles in
permutations of the set~$\{1,\dots,n\}$.

\begin{theoremMy} {\rm \cite{SAI}}
$|\bar{\mathcal{\chi}}_n|/|{\mathcal{\chi}}_n| \longrightarrow 1$
as $n\to \infty$.
\end{theoremMy}

{\bf Proof.} Let $\mathcal{S}_n$~be  the set of all permutations
of the set~$\{1,\dots,n\}$. Consider a random permutation $s$ from
$\mathcal{S}_n$. By $\xi(s)$ denote the number of cycles in
permutation~$s$. It is known (see e.g.~\cite{Feller70}) that the
expectation $E[\xi(s)]$ of random variable $\xi(s)$ is equal to
$\sum\limits_{i=1}^n\frac{1}{i}$ and the variance $Var[\xi(s)]$
equals  $\sum\limits_{i=1}^n\frac{i-1}{i^2}$. Let
$\bar{\mathcal{S}}_n$~denote the set of permutations
from~$\mathcal{S}_n$, where the number of cycles is at
most~${1.1~{\rm ln}(l)}$. Then, using Chebychev's
inequality~\cite{Feller70}, we get

\begin{equation}|\bar{\mathcal{S}}_n|/|{\mathcal{S}}_n| \longrightarrow 1\ \mathrm{as} \ n\to \infty. \label{AlAll1}\end{equation}

Now let $\mathcal{S}'_n$ denote the set of permutations from~$\mathcal{S}_n$,
which do not contain the cycles of length~1, and let $\mathcal{S}^{(i)}_n$ be the set of permutations from~$\mathcal{S}_n$,
which contain a cycle  with element $i,\ i=1,\dots,n$. Using the principle of inclusion and exclusion~\cite{Riordan58}, we obtain

$$\left|\mathcal{S}_n\setminus \mathcal{S}'_n\right|=\left|\bigcup_{1\le i \le n} \mathcal{S}^{(i)}_n\right|=
\sum_{i=1}^{n}\left|\mathcal{S}^{(i)}_n\right|-\sum_{1\le i\ne j\le n}\left|\mathcal{S}^{(i)}_n \bigcap \mathcal{S}^{(j)}_n\right|+$$
$$\sum_{1\le i\ne j\ne k \le n}\left|\mathcal{S}^{(i)}_n \bigcap \mathcal{S}^{(j)}_n  \bigcap \mathcal{S}^{(k)}_n\right|-\dots=
n!-C_n^2(n-2)!+C_n^3(n-3)!-\dots\le $$
$$ \frac{n!}{2}+ \frac{n!}{6}=\frac{2}{3}n!=\frac{2}{3}|\mathcal{S}_n|.$$

Therefore, \begin{equation}|\mathcal{S}'_n|\ge \frac{1}{3}|\mathcal{S}_n|. \label{AlAll2}\end{equation}

Combining (\ref{AlAll1}) and (\ref{AlAll2}), we get

\begin{equation}|\bar{\mathcal{S}}'_n|/ |\mathcal{S}'_n| = 1- \frac{|\mathcal{S}'_n\backslash \bar{\mathcal{S}}'_n|}{|\mathcal{S}'_n|}
\ge 1- \frac{3|\mathcal{S}'_n\backslash \bar{\mathcal{S}}'_n|}{|\mathcal{S}_n|}\ge
1- \frac{3|\mathcal{S}_n\backslash \bar{\mathcal{S}}_n|}{|\mathcal{S}_n|} \mathop{\longrightarrow }\limits_{n\to +\infty} 1, \label{AlAll3}\end{equation}
where $\bar{\mathcal{S}}'_n={\mathcal{S}}'_n \cap \bar{\mathcal{S}}_n$.

The values~$|\bar{\mathcal{\chi}}_n|$ and $|{\mathcal{\chi}}_n \backslash \bar{\mathcal{\chi}}_n|$ may be bounded, using the following approach.
We assign any permutation~$s \in \mathcal{S}'_l$, $l\le n$, a set of
bipartite graphs~${{\mathcal{\chi}}_n (s)\subset {\mathcal{\chi}}_n}$ as follows.
First of all let us assign an arbitrary set of $n-l$ edges to be
special. Then the non-special vertices $\{i_1,i_2,\dots,i_l\}\subset
X_n$ of the left-hand part, where $i_j<i_{j+1},\ j=1,\dots,l-1$,
are now partitioned into $\xi(s)$ blocks, where $\xi(s)$
is the number of cycles in permutation~$s$. Every cycle
$(t_1,t_2,\dots,t_r)$ in permutation~$s$ corresponds to some
sequence of vertices with indices
$\{i_{t_1},i_{t_2},\dots,i_{t_r}\}$ 
belonging to the block associated with this cycle. Finally, it is
ensured that for each pair of vertices $\{i_{t_j},\
i_{t_{j+1}}\}$, $j=1,\dots,r-1$, as well as for the pair
$\{i_{t_{r}}, \ i_{t_1}\},$ there exists a vertex in the
right-hand part~$X$ which is adjacent to both vertices of the pair.
Except for special edges and blocks additional edges are allowed in graphs from class ${{\mathcal{\chi}}_n (s)}$.
These edges are adjacent to the special vertices of the left-hand part such that the degree
of any vertex of the left-hand part is not greater than two. Moreover, additional edges
should not lead to creating  new blocks.

There are~$n!$ ways to associate vertices of the left-hand part to
vertices of the right-hand part, therefore the number of different
graphs from class~${{\mathcal{\chi}}_n (s)}$, $s \in \mathcal{S}'_l$,
$l\le n$, is
$|{{\mathcal{\chi}}_n (s)}|=C_n^l\frac{n!}{2^{\xi_1(s)}}h(n,l)$, where
function $h(n,l)$ depends only on $n$ and $l$, and  $\xi_1(s)$ is the number of cycles of length two in
permutation~$s$. Division by~$2^{\xi_1(s)}$  is here due
to the fact that for each block that corresponds to a cycle of
length two in~$s$, there are two equivalent ways to number
the vertices in its right-hand part.

Let~${s=c_1c_2\dots c_{\xi(s)}}$ be a permutation from
set~$\mathcal{S}'_l$, represented by cycles~$c_i$,
${i=1,\dots,\xi(s)},$ and let $c_j$ be an arbitrary cycle of
permutation~$s$ of length at least three, $j=1,\dots,\xi(s)$. Permutation~$s$ may be transformed
into permutation~$s^1$,

\begin{equation}
s^1=c_1c_2\dots c_{j-1} c_j^{-1}c_{j+1}\dots
c_{\xi(s),}\label{cicle}
\end{equation}
by reversing the cycle~$c_j$. Clearly, permutation~$s^1$
induces the same subset of bipartite graphs in class~${\mathcal{\chi}}_n$ as the
permutation~$s$ does. Thus any two permutations~$s^1$
and $s^2$ from set~$\mathcal{S}'_l$, $l\le n$, induce the same
subset of graphs in~${\mathcal{\chi}}_n$, if one of these permutations may be
obtained from the other one by several transformations of the
form~(\ref{cicle}). Otherwise the two induced subsets of graphs do
not intersect. Besides  that ${{\mathcal{\chi}}_n (s^1)}\cap
{{\mathcal{\chi}}_n (s^2)}=\emptyset$ if $s^1\in \mathcal{S}'_{l_{1}}$,
$s^2\in \mathcal{S}'_{l_{2}}$, $l_1\ne l_2$.

On one hand, if $s\in \bar{\mathcal{S}}'_l$, $l\le n$,  then
${\mathcal{\chi}}_n(s)\subseteq \bar{\mathcal{\chi}}_n $. On the other hand, if
${s\in \tilde{\mathcal{S}}'_l}:=\mathcal{S}'_l \backslash \bar{\mathcal{S}}'_l $, $l< n$, then either
${\mathcal{\chi}}_n(s)\subseteq \bar{\mathcal{\chi}}_n$ or, alternatively,
${{\mathcal{\chi}}_n(s)\subseteq {\mathcal{\chi}}_n\backslash \bar{\mathcal{\chi}}_n}$ may hold. Therefore,

\begin{equation}
|\bar{\mathcal{\chi}}_n|\ge\sum_{l=2}^n\sum_{s \in
\bar{\mathcal{S}}'_l}C_n^l\frac{n!}{2^{\xi_1(s)}2^{\xi(s)-\xi_1(s)}}h(n,l)=\sum_{l=2}^n\sum_{s\in \bar{\mathcal{S}}'_l}C_n^l\frac{n!}{2^{\xi(s)}}h(n,l)\ge \label{eq1}
\end{equation}

$$
\ge \sum_{l=2}^n |\bar{\mathcal{S}}'_l |\cdot C_n^l\frac{n!}{2^{1.1\mathrm{ln}(l)}}h(n,l)\ge \sum_{l=\lfloor 1.1{\rm
ln}(n)\rfloor}^n |\bar{\mathcal{S}}'_l |\cdot C_n^l\frac{n!}{2^{1.1\mathrm{ln}(l)}}h(n,l),
$$

\begin{equation}
|{\mathcal{\chi}}_n\backslash \bar{\mathcal{\chi}}_n|\le \sum_{l=\lfloor
1.1{\rm ln}(n)\rfloor}^n\sum_{s
\in \tilde{\mathcal{S}}'_l}C_n^l\frac{n!}{2^{\xi(s)}}\le \sum_{l=\lfloor 1.1{\rm
ln}(n)\rfloor}^n |\tilde{\mathcal{S}}'_l |\cdot C_n^l\frac{n!}{2^{1.1\mathrm{ln}(l)}}h(n,l).
\label{eq2}
\end{equation}

Now assuming $\psi(n)=\max\limits_{l=\lfloor 1.1{\rm
ln}(n)\rfloor,\dots,n}|\tilde{\mathcal{S}}'_{l}|/|\bar{\mathcal{S}}'_{l}|$ and taking
into account~(\ref{eq1}),~(\ref{eq2}) and (\ref{AlAll3}),
we obtain

\begin{equation}
\frac{|{\mathcal{\chi}}_n\backslash \bar{\mathcal{\chi}}_n|} {|\bar{\mathcal{\chi}}_n|}\le \psi(n) \to 0 \mbox{ as }
n\to +\infty. \label{eq5}
\end{equation}
Finally, the statement of the theorem follows from~(\ref{eq5}). Q.E.D.
\end{document}